# Ferroelectric Fluids for Nonlinear Photonics: Evaluation of Temperature Dependence of Second-Order Susceptibilities

M. Lovšin[1,2], L. Cmok[1], C. J. Gibb[3], J. Hobbs[4], R. J. Mandle[3,4], A. Mertelj[1], I. Drevenšek-Olenik[1,2], N. Sebastian[1]

1 Jožef Stefan Institute, Ljubljana, Slovenia
2 University of Ljubljana, Faculty of Mathematics and Physics, Ljubljana, Slovenia
3 School of Chemistry, University of Leeds, Leeds, UK
4 School of Physics and Astronomy, University of Leeds, Leeds, UK

**Abstract:** Ferroelectric nematic fluids are promising materials for tunable nonlinear photonics, with applications ranging from second harmonic generation to sources of entangled photons. However, the few reported values of second-order susceptibilities vary widely depending on the molecular architecture. Here, we systematically measure second-order NLO susceptibilities of five different materials that exhibit the ferroelectric nematic phase, as well as the more recently discovered layered smectic A ferroelectric phase. The materials investigated include archetypal molecular architectures as well as mixtures showing room-temperature ferroelectric phases. The measured values, which range from 0.3 to 20 pm/V, are here reasonably predicted by combining calculations of molecular-level hyperpolarizabilities and a simple nematic potential, highlighting the opportunities of modelling-assisted design for enhanced NLO ferroelectric fluids.

## 1. Introduction

Nonlinear optical (NLO) materials play a crucial role in a wide range of applications, from frequency conversion to laser systems, telecommunications, and even quantum technologies [1–3]. Historically, commercial NLO components have relied on solid-state crystals [4], like lithium niobate (LN), potassium dihydrogen phosphate (KDP), potassium titanyl phosphate (KTP), and barium borate (BBO), which all exhibit good NLO properties yet lack the tunability and flexibility of soft matter materials.

A new alternative emerged in 2017 with the discovery of the first two ferroelectric nematic liquid crystalline (FNLC) materials, RM734 and DIO [5–9]. These materials exhibit a nematic liquid crystal phase with a macroscopic electric polarization: the ferroelectric nematic ($N_F$) phase. Their polar order breaks the inversion symmetry, enabling second-order NLO processes, like optical second harmonic generation (SHG), to emerge. To achieve efficient SHG, two material properties are important: large NLO susceptibility coefficients and the possibility to attain material orientation/configuration that leads to phase matching (PM). The second-order NLO susceptibility is the third-order tensor that connects the second-order electric polarization to the electric field: $\vec{P} = \epsilon_0 \chi^{(2)} : \vec{E}\vec{E}$. By convention, the following NLO tensor is introduced: $d_{ijk} = \frac{1}{2}\chi^{(2)}_{ijk}$. Considering permutation symmetries, the last 2 indices can be replaced by one, resulting in a second-order tensor with 18 components: $d_{ijk} \rightarrow d_{il}$. The first measurements of the NLO coefficients report the values $d_{33} = 5.6 \frac{pm}{V}$ for RM734 [10] and $d_{33} = 0.24 \frac{pm}{V}$ for DIO [11]. Although these values are about one order of magnitude lower than those of traditional NLO crystals (e.g., LN: $d_{33} = 25.7 \frac{pm}{V}$ [12]) and on par with quartz ($d_{11} = 0.3 \frac{pm}{V}$ [12]), targeted molecular design focused on enhancing hyperpolarizability along the donor-acceptor axis may possibly lead to even surpassing them. For instance, adding chromophores to RM734 led to an increase in the $d_{33}$ up to 25 $\frac{pm}{V}$ [13] in the absorption regime. Another material requirement for the efficient SHG process is the possibility to obtain phase matching between the fundamental and the second harmonic optical waves. In this respect, FNLCs provide several new options. For instance, conversion efficiency can be significantly increased by introducing chiral dopants that induce a periodic helielectric nematic structure and subsequently enable quasi-phase matching conditions to be realized [14–18].

Still, the greatest advantage of FNLCs against NLO crystals is in novel capabilities to control the material orientation and therefore its effective NLO response by various passive as well as active methods. It has been shown that targeted surface alignment allows spatial control of the polarization direction

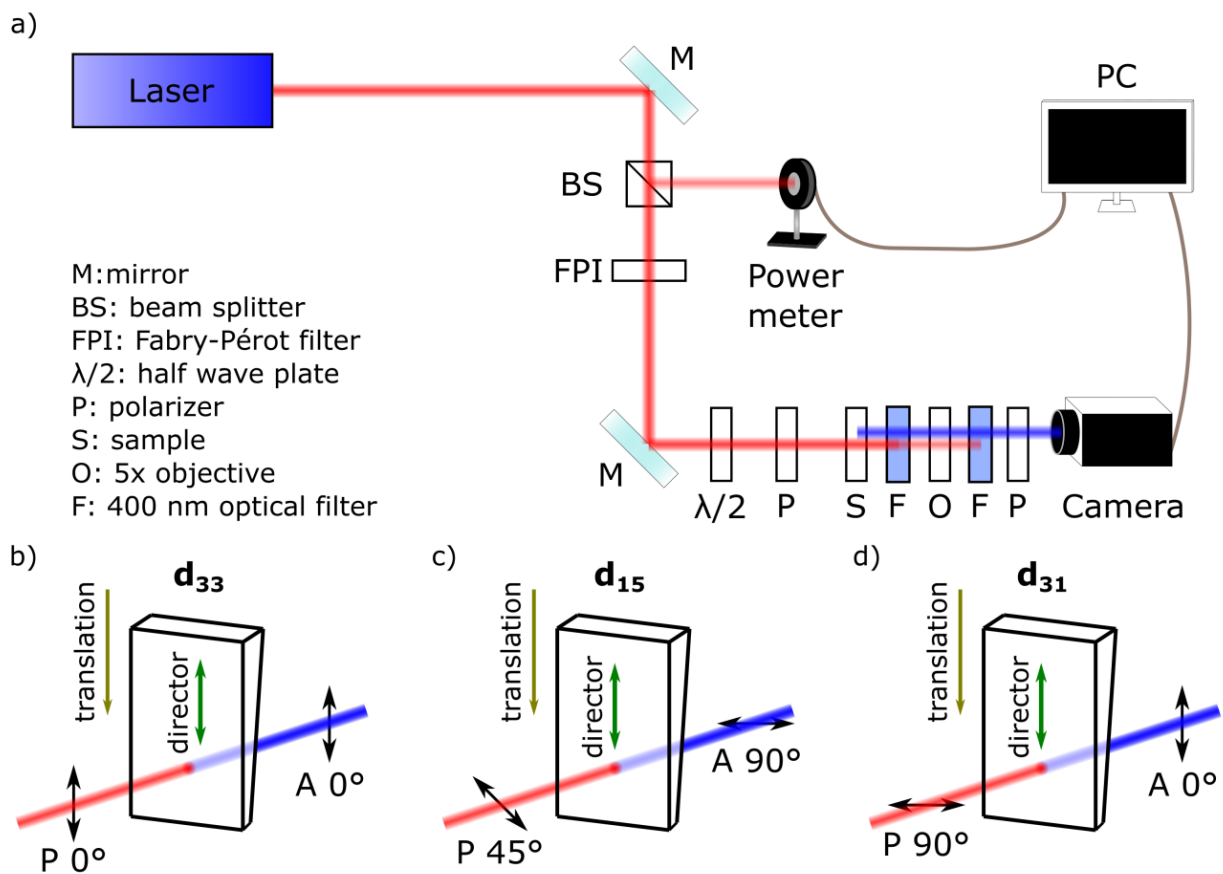

**Figure 1.** SHG setup. a) Scheme of the SHG Maker Fringes measurement system as described in Methods. Orientations of the fundamental beam polarization and the analyzer relative to the director (green arrow) for the measurement of b) $d_{33}$, c) $d_{15}$, and d) $d_{31}$.

[19] and enables the creation of 2D SHG patterns on a micrometer scale [20]. Another strategy is to actuate the FNLC orientation using external electric fields [21], which leads to strong modulation of the SHG signal. In addition to the SHG process, another promising applicative avenue is electric-field tunable spontaneous parametric down-conversion (SPDC), which represents a pathway for designing tunable sources of entangled photon pairs from FNLCs [22].

The discovery of the ferroelectric nematic phase ($N_F$) was followed by the discovery of many new phases, like the heliconical ferroelectric nematic phase, which exhibits polar order and chirality without any chiral dopants [23]. Another phase, interesting for the NLO applications, is a polar smectic phase ($SmA_F$) [24–26]. This phase also exhibits NLO properties as its symmetry is also broken by polar order and has the same $C_{\infty v}$ symmetry as the $N_F$ phase. Besides these, phases like the ferroelectric twist-bent nematic phase ($N_{TBF}$) [23], and polar chiral tilted smectic phase ($SmC_P^H$) [24], have also been reported.

Although the initial assessment of the SHG properties of RM734 and DIO has been conducted, the thorough characterization of the nonlinear susceptibility tensor coefficients ($d_{ij}$) and systematic measurements of dispersion of refractive indices are needed before planning any NLO components. In this work, we employ the Maker fringes method to perform the first temperature dependence analysis of the $d_{ij}$ coefficients of two standard FNLC materials, RM734 and DIO. We also demonstrate the analysis of the $d_{ij}$ coefficients of two materials with room-temperature $N_F$ phase, FNLC-1571 and F7, as well as the first comparative study of SHG between 2 pure compounds and their mixture. Furthermore, we report the first characterization of the SHG process in the $SmA_F$ and $SmC_P^H$ phases. The molecular structures of the investigated materials are shown in Figure S1. In the end, we present a computational method that could guide the design of new polar liquid crystals with optimal NLO properties.

## 2. Results

We measured $d_{ij}$ coefficients using Maker fringes method and employing wedge cells as described in Materials and Methods. The LC phases $N_F$ and $SmA_F$ have $C_{\infty v}$ symmetry, and the symmetry of the SmC$_P^H$ can be treated as C$_{\infty v}$ as well, because the period of the helical structure is much smaller than the optical wavelengths. Consequently, the number of independent nonzero components of the NLO susceptibility for all the investigated LC phases reduces to 3:

$$d = \begin{pmatrix} 0 & 0 & 0 & 0 & d_{15} & 0 \\ 0 & 0 & 0 & d_{15} & 0 & 0 \\ d_{31} & d_{31} & d_{33} & 0 & 0 & 0 \end{pmatrix} \quad (1)$$

where the z-axis is along the director. According to Kleinman symmetry, in the absence of dispersion and absorption in the medium, the $d_{31}$ should be equal to $d_{15}$, reducing the number of distinct components to 2. However, liquid crystals are dispersive materials, and the components remain distinct, although their values are expected to be similar.

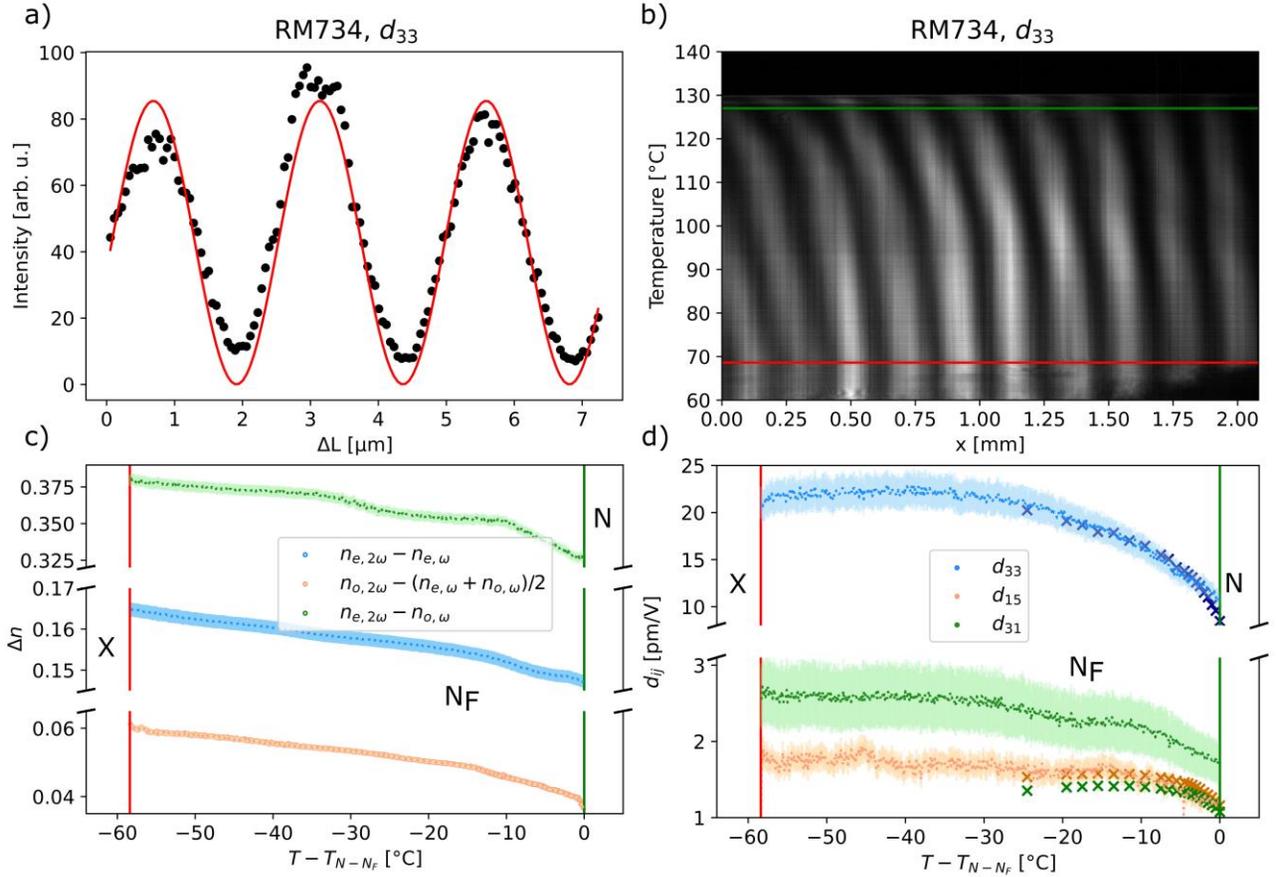

**Figure 2.** Second-order susceptibility measurement of RM734. a) SHG intensity as a function of cell thickness for RM734 (the measurement of $d_{33}$). b) The temperature dependence of the SHG intensity profile (the measurement of $d_{33}$). c) The measured refractive index mismatch $\Delta n$ and d) the corresponding NLO susceptibility components $d_{33}$, $d_{15}$, and $d_{31}$ of RM734. The measured data (dots) with their uncertainties (shaded area) compared to the values calculated from molecular hyperpolarizability (crosses). The first SHG signal was detected at 130.5°C, and the $N_F$ phase stabilized at 127°C (green line). The SHG signal disappeared at 68.6°C (red line).

The first material we measured was RM734. It aligned well in the commercial wedge cell, which has a sufficient slope to allow for the imaging of multiple Maker fringes within the beam spot. In order to determine all three components of the nonlinear susceptibility tensor, the three geometries described in the Methods were explored. The results are summarized in Figure 2. The component with the largest value was $d_{33}$, which, with decreasing temperature, increased in value up to $22.5 \pm 2$ pm/V.

The maximum measured value of $d_{15}$ of RM734 is an order of magnitude lower than $d_{33}$, reaching $d_{15} = 1.8 \pm 0.2$ pm/V. For all measured materials, the refractive index mismatch $\Delta n_{31} = n_{e,2\omega} - n_{o,\omega}$ is expected to be significantly larger than $\Delta n_{15} = n_{o,2\omega} - (n_{e,\omega} + n_{o,\omega})/2$ ( $n_e$ denotes the extraordinary refractive index, oriented along the director, and $n_o$ the ordinary refractive index, oriented perpendicular to the director). Consequently, as the SHG intensity is inversely proportional to $\Delta n^2$ (Eq. 9), it follows that even when $d_{31}$ is comparable to $d_{15}$, the amplitude of the Maker fringes associated with $d_{31}$ is much lower, making this component more difficult to measure. This is an issue that will also appear and affect the rest of the studied materials. In our analysis, RM734 was the only material for which the $d_{31}$ component could be measured. For all other measured materials, the maximum signal in the measurements of $d_{31}$ was too low to be acquired with the current setup.

We then checked DIO, for which there is also some published data on the NLO susceptibility [11]. DIO did not align well in the commercial cell, and a custom-made cell was needed. This cell has a lower wedge slope, and it had to be translated along the wedge in order to detect enough Maker fringes. We were able to measure the $d_{33}$ and $d_{15}$ components (Figure 3), but not the $d_{31}$. The reason for this could be a small period of the Maker fringes or a small NLO coefficient, as both of those cause a low maximum SHG signal of the fringes. The obtained values of both $d_{33}$ and $d_{15}$ are similar to each other and much lower than those of RM734, around 0.35 pm/V. In the Supplementary Information of [19], it was already shown that, in contrast to RM734, in DIO, the $d_{15}$ component significantly contributes to the net SHG signal. The authors reported that the SHG signal was stronger when the pump beam polarization was at 45° with respect to the director than when it was parallel to it, which indicates that the cell thickness used in their experiment was closer to the Maker minimum for $d_{33}$ than for $d_{15}$. DIO is a fundamental part of material F7, which also contains C1.

The polar smectic material C1 was measured both in the SmA$_F$ and the SmC$_P^H$ phases. The maximal measured values of $d_{33}$ and $d_{15}$ in the SmA$_F$ phase are $d_{33} = 3.9 \pm 0.6$ pm/V and $d_{15} = 0.35 \pm 0.06$ pm/V at 88°C, which is the temperature of the phase transition to SmC$_P^H$. The gap in Figure 4 corresponds to the range of temperatures during SmA$_F$ – SmC$_P^H$ phase transition, in which the SmC$_P^H$ phase has not stabilized yet. As mentioned earlier, since the helical pitch is much smaller than the optical wavelengths, the symmetry of the SmC$_P^H$ phase can be treated as C$_{\infty v}$. The effective values of $d_{33}$ and $d_{15}$ are somewhat smaller than before the transition, and they decrease with cooling. As in DIO, also in C1, the $d_{31}$ component could not be measured because the SHG signal was too low.

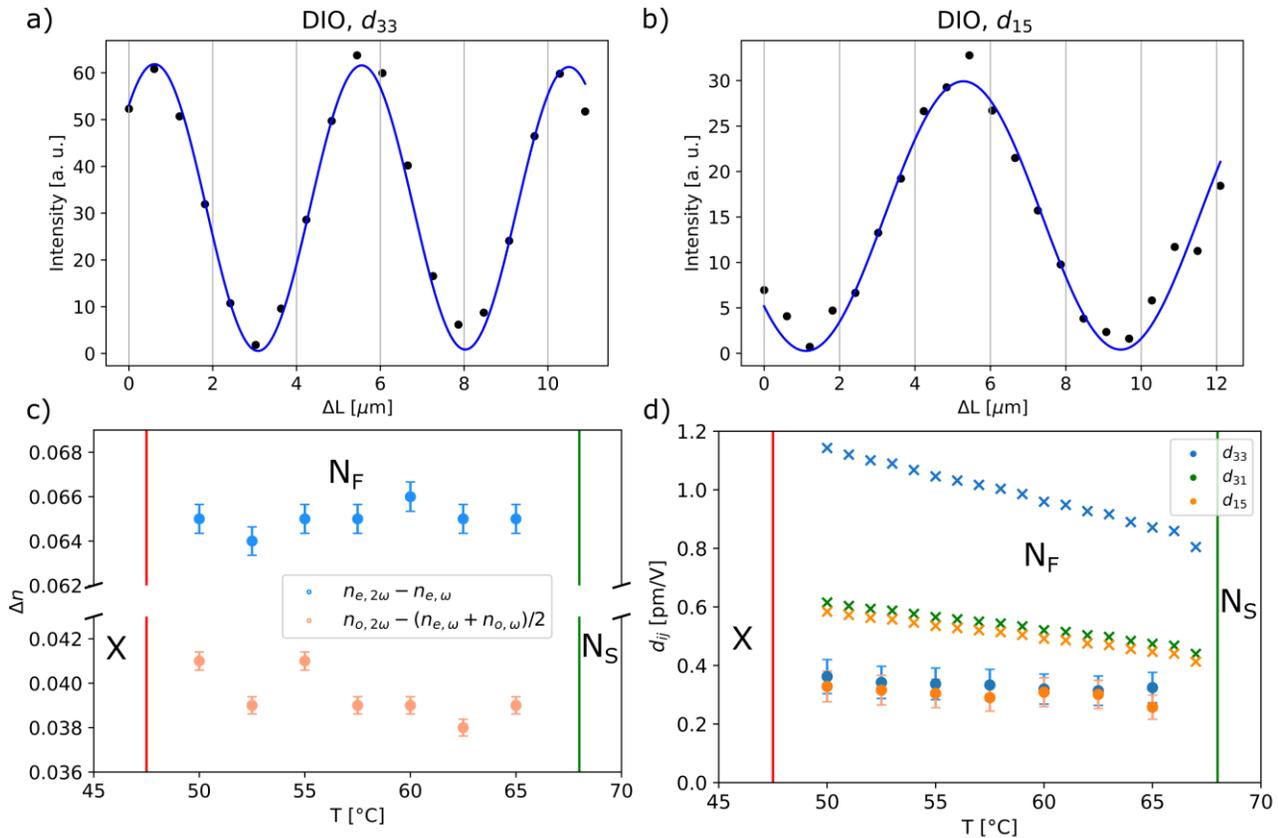

**Figure 3.** Second-order susceptibility measurement of DIO. SHG intensity as a function of cell thickness for DIO (the measurement of a) $d_{33}$ and b) $d_{15}$). c) The measured refractive index mismatch Δn and d) the corresponding NLO susceptibility components $d_{33}$ and $d_{15}$ of DIO. The measured data (dots) with their error bars compared to the values calculated from molecular hyperpolarizability (crosses). The phase transition N$_S$-N$_F$ occurred at 68°C (green line), and the SHG signal disappeared at 47.5°C (red line).

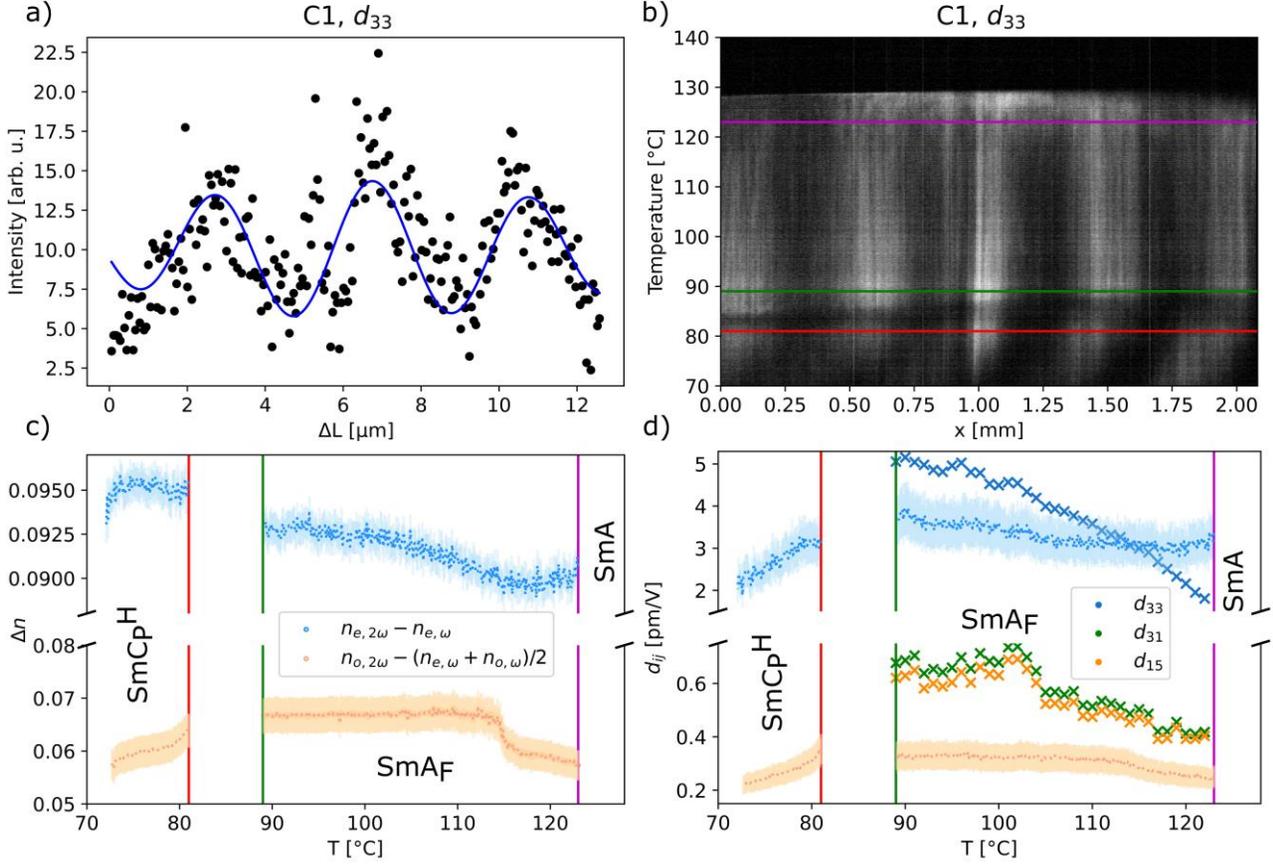

**Figure 4.** Second-order susceptibility measurement of C1. a) SHG intensity as a function of cell thickness for C1 (the measurement of $d_{33}$). b) The temperature dependence of the SHG intensity profile (the measurement of $d_{33}$). c) The measured refractive index mismatch Δn and d) the corresponding NLO susceptibility components $d_{33}$ and $d_{15}$ of C1. The measured data (dots) with their uncertainties (shaded area) compared to the values calculated from molecular hyperpolarizability (crosses). The phase transition SmA-SmA$_F$ of C1 occurred at 123°C (magenta line), the phase transition to SmC$_P^H$ started at 89°C (green line), and the SmC$_P^H$ phase stabilized at 81°C (red line).

The last two materials that we investigated are especially promising for the NLO applications, given that they exhibit the N$_F$ phase at room temperature. For the material F7, which is a mixture of 30% C1 and 70% DIO, we expected values of the $d_{ij}$ between those measured for the pure materials. The obtained peak values are $d_{33} = 2.1 \pm 0.2 \, \frac{pm}{V}$ and $d_{15} = 0.47 \pm 0.07 \, \frac{pm}{V}$ (Figure 5), while the weighted average of C1 and DIO would be $d_{33} = 1.4 \, \frac{pm}{V}$ and $d_{15} = 0.33 \, \frac{pm}{V}$. As in previous cases, the signal of the $d_{31}$ was too low to be measured.

The last investigated material, ferroelectric nematic material FNLC-1571, exhibits the N$_F$ phase at room temperature as well and is now widely used as a study material. In a previous study, a value of $d_{33} \approx 20$ pm/V has been reported for $\lambda = 1370$ nm, determined by a measurement using single sample thickness and compared with that of a reference sample [22]. As the FNLC-1571 did not align well in the commercial cell, we used two different custom-made wedge cells, one with a wedge slope $\alpha = 1.27°$ and one with a wedge slope $\alpha = 0.07°$, with the result of the latter allowing us to associate the non-zero signal of fringes' minima (Figure 6.a and Figure S10) to the large wedge slope of the former. The temperature dependence for the different coefficients and involved dispersions on heating is shown in Figure 6. The measurements of $d_{33}$ and $d_{15}$ show weak temperature dependence (for $d_{33}$ from 9 to 11 pm/V and for $d_{15}$ around 0.45 pm/V). A slight difference in the transition temperature into the higher temperature antiferroelectric phase between the measurements of $d_{33}$ and $d_{15}$ was detected, most probably provoked by the increased laser power and consequent cell heating, in the $d_{15}$ measurements performed without the Fabry-Pérot filter. Unfortunately, no signal for $d_{31}$ was observed, which could be attributed to a large refractive index mismatch. However, as for the rest of the materials, it is expected to have a similar value to $d_{15}$.

## 3. Discussion

We have measured the NLO susceptibility of several materials, and the results show that they vary significantly from one material to another (from $d_{33} = 0.36 \pm 0.05 \frac{pm}{V}$ in DIO to $d_{33} = 22.5 \pm 2 \frac{pm}{V}$ in RM734). The difference stems from their molecular structure, highlighting the challenge of designing new molecular structures that will result in optimal NLO properties while preserving the $N_F$ phase. While a comprehensive relationship between NLO properties and molecular structure is outside the scope of this work, due to the limited number of materials available for study, the large NLO susceptibility of RM734 probably stems from the push-pull system established by the 2,4-dimethoxybenzoate (donor) and 4-nitrophenyl (acceptor) structure. For DIO, C1, and F7, the donor (5-propyl-1,3-dioxane) and acceptor (3,4,5-trifluorobenzene) are far weaker, manifesting in smaller NLO susceptibilities.

With the growing number of molecules exhibiting polar phases being designed, it would be beneficial to be able to calculate their potential NLO susceptibility values. To achieve this, it is necessary to relate the measured NLO susceptibility to the molecular properties of the analyzed materials. According to the oriented gas model, the NLO tensor components can be calculated as [10]:

$$d_{ijk} = N F_\omega F_\omega F_{2\omega} \langle \beta_{ijk} \rangle, \quad (2)$$

where N is the volume density of the material, $N = \frac{N_A \rho}{M}$, $N_A$ is Avogadro's constant, $\rho$ is density, and M is the molar mass of the material. $F_\alpha$ are the local field factors for the fundamental and the second harmonic frequencies, and $\langle \beta_{ijk} \rangle$ is the thermally averaged first molecular hyperpolarizability.

The local field factor is given by

$$F(\omega) = \frac{n^2 + 2}{3}. \quad (3)$$

Usually, it is calculated by taking the average $n = \frac{1}{3}(2n_{o,\omega} + n_{e,\omega})$. Considering dispersion, and calculating for each refractive index separately, results in approximately 20% higher local field correction factor in our case (Fig. S8). Additional improvement of the local field factor calculation accuracy can be done if the uniaxial liquid crystals are treated as anisotropic crystals with a tetragonal lattice approximation [27]:

$$\begin{aligned} F(\omega_e) &= 1 - L_{ee}(n_e^2 - 1), \\ F(\omega_o) &= 1 - L_{oo}(n_o^2 - 1), \end{aligned} \quad (4)$$

where $L_{ee}$ and $L_{oo}$ are the diagonal components of the Lorentz factor tensor, which depend on the degree of anisotropy. As the tensor is traceless, the relation between its components is $L_{oo} = \frac{1}{2}(1 - L_{ee})$. The $L_{ee}$ was determined to be between 0.2 and 0.28 for a few typical nematic liquid crystals [28]. We assume the FNLCs to be in the same order of magnitude.

Another way to describe the local field factor has been proposed to introduce anisotropy in the model for isotropic liquids [29]:

$$\begin{aligned} F(\omega_e) &= \frac{n_e^2 + 2}{3} + \eta_{ee}(n_e^2 - 1), \\ F(\omega_o) &= \frac{n_o^2 + 2}{3} + \eta_{oo}(n_o^2 - 1), \end{aligned} \quad (5)$$

where $\eta_{ee}$ and $\eta_{oo}$ are components of the anisotropy tensor that depend on the pair distribution function. The values for a typical nematic liquid crystal were determined to be between 0.06 and 0.08 [29].

To assess the effect of correction, we plot the local field factor dependence on the level of anisotropy (Fig. S8). We see that the values of $F(\omega)$ calculated in this way are lower, and the calculation with the average $n = \frac{1}{3}(2n_{o,\omega} + n_{e,\omega})$ gives us a good approximation in the highlighted grey area.

For calculating the thermally averaged first molecular hyperpolarizability $\langle \beta_{ijk} \rangle$, the corresponding thermal averages for the polar angle are needed. They can be calculated from the nematic potential, which we take as proposed in Folcia et al. [10]:

$$U(\theta) = A(T) \sin^2 \theta + B(T) \sin^2 \frac{\theta}{2}, \quad (6)$$

|  | RM734 ($N_F$, 93°C) | DIO ($N_F$, 47°C) | FNLC-1571 ($N_F$, 23°C) | C1 (SmA$_F$, 88°C) | F7 ($N_F$, 40°C) |
|---|---|---|---|---|---|
| $d_{33}$ [pm/V] | 22.5 ± 2 | 0.36 ± 0.05 | 11.0 ± 0.9 | 3.9 ± 0.6 | 2.1 ± 0.2 |
| $d_{15}$ [pm/V] | 1.8 ± 0.2 | 0.32 ± 0.05 | 0.46 ± 0.07 | 0.35 ± 0.06 | 0.47 ± 0.07 |
| $d_{31}$ [pm/V] | 2.7 ± 0.4 |  |  |  |  |

**Table 1.** Maximal values of measured second-order NLO susceptibility components.

where $\theta$ is the polar angle and parameters $A(T)$ and $B(T)$ account for the usual nematic potential and the new contribution due to polarity, and can be calculated from the measured data of polarization and birefringence of the material

$$\langle \cos\theta \rangle = \frac{P(T)}{N\mu}, \quad (7)$$

and

$$\frac{S}{S_{NP}} = \frac{\Delta n}{\Delta n_{NP}} \quad (8)$$

where $P(T)$ is the spontaneous polarization of the material, $\mu$ is the dipole moment of the molecule, and $S_{NP}$ and $\Delta n_{NP}$ are the extrapolated order parameter and the birefringence in the case of $B(T) = 0$ (standard nematic phase). The values of $\Delta n_{NP}$ can be obtained by extrapolating the birefringence of the N phase into the N$_F$ phase temperature range. As the birefringence in the SmA phase mainly comes from the orientational order, not from the translational one, Equations (6)-(8) can also be applied for the SmA–SmA$_F$ phase transition in C1. As the dipole moment $\mu$ is not parallel to the molecular long axis, only its component along the long axis was considered in the calculations. All other data used for calculations are presented in Table S4.

The thermally averaged hyperpolarizability components in the laboratory frame $\langle \beta_{ijk} \rangle$ were calculated from the ones in the molecular frame according to Equations S1-S3 [30,31]. The components of the hyperpolarizability tensor in the molecular frame were obtained by calculating the molecular electronic structure with the Gaussian software package at the M06HF-D3/aug-cc-pVTZ level at a frequency corresponding to 800 nm. Values thus obtained are given in Tables S1-S3. However, it should be noted here that Gaussian calculations are for a single molecular conformation and thus limit the accuracy of the results.

As a case example, we start with RM734, for which the polarization data were taken from [8] and the measured birefringence at 405 nm is shown in Fig. S9. The calculated NLO tensor components are shown in Fig. 2 together with the experimental data, showing a very good match with the measurements for $d_{33}$ and $d_{15}$, while a slightly worse match for $d_{31}$, likely due to the fact that $d_{31}$ was measured without the Fabry-Pérot filter. However, it is worth recalling here that several approximations are needed for these calculations, especially in regards of local fields, so even a larger mismatch for any of the components would have been a positive result.

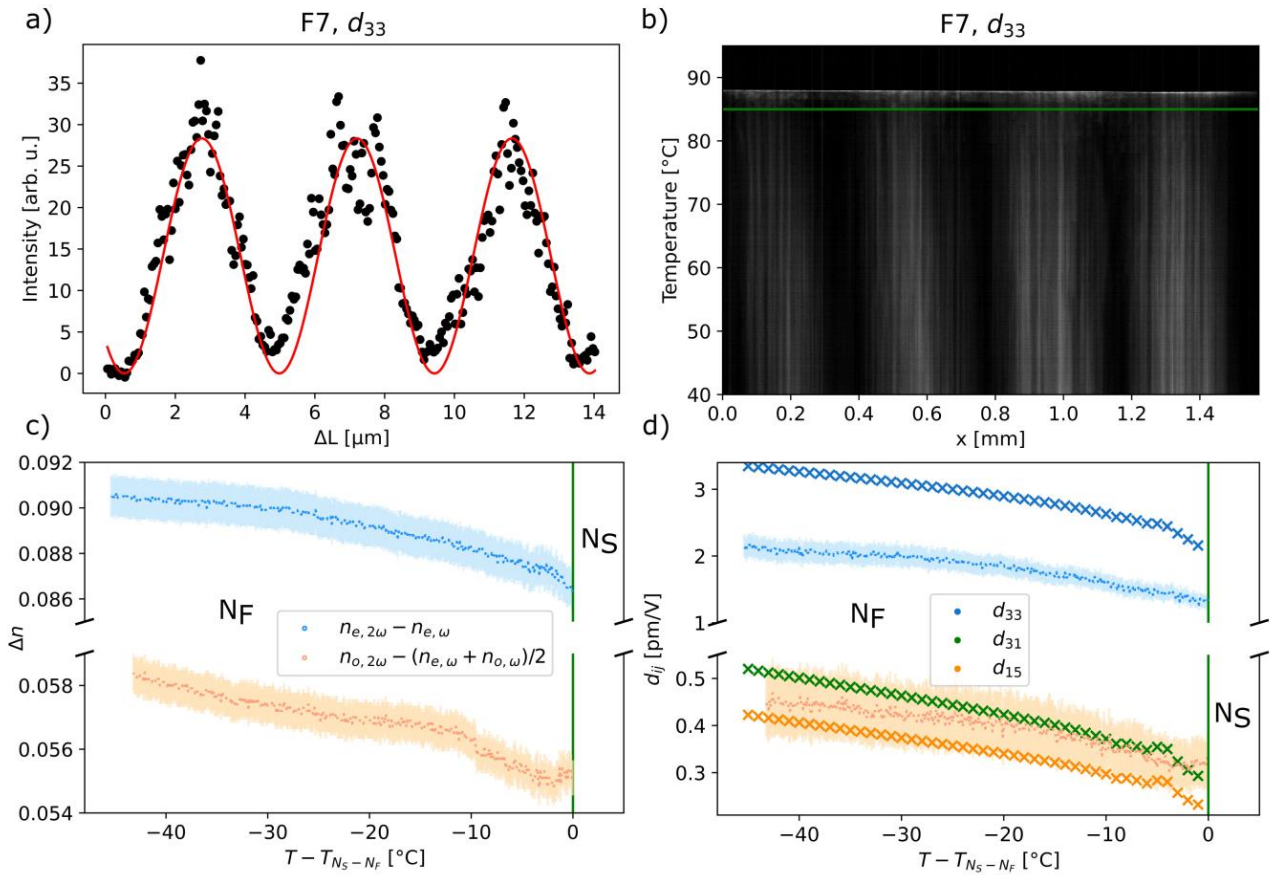

**Figure 5.** Second-order susceptibility measurement of F7. a) SHG intensity as a function of cell thickness for F7 (the measurement of d$_{33}$). b) The temperature dependence of the SHG intensity profile (the measurement of d$_{33}$). c) The measured refractive index mismatch Δn and d) the corresponding NLO susceptibility components d$_{33}$ and d$_{15}$ of F7. The measured data (dots) with their uncertainties (shaded area) compared to the values calculated from molecular hyperpolarizability (crosses). The phase transition N$_S$-N$_F$ of F7 occurred at 85°C (green line).

We compare our results with the only available data for $d_{33}$ of RM734 at a single temperature [10]. For that, we follow their procedure of calculating the hyperpolarizability tensor only from the reported measured values for the NO$_2$-π-O group and use the fundamental wavelength of 800 nm (as opposed to 1064 nm in their measurement). In their case, the hyperpolarizability is calculated with the two-level dispersion model, assuming the resonance wavelength at $\lambda_{max} = 304$ nm and data from [32]: $\beta_{1907nm} = 3.0 * 10^{-30}$ esu, leading to: $\beta_{1064nm} = 4.2 * 10^{-30}$ esu and equivalently for our case $\beta_{800nm} = 7.3 * 10^{-30}$ esu. Finally, the value calculated in this way of d$_{33}$ at 1064 nm is 6.9 pm/V, and at 800 nm it is 12.0 pm/V. At 118°C, Folcia et al. reported a measured value of $d_{33} = 5.6\ pm/V$, while our measurement at the same temperature gives $d_{33} = 14.9 \pm 2\ pm/V$, highlighting the measurements presented here are consistent with those reported for a single temperature by Folcia et al. Altogether, this highlights the important dependence on wavelength for the second-order NLO susceptibilities of these materials in the 800-1200 nm range.

We repeat the same procedure for DIO, C1, and F7. The measured polarization and birefringence measurement data are shown in Fig. S9. The polarization of C1 and F7 was determined via the standard triangular wave method, while the polarization data for DIO were taken from [5]. The molecular parameters of F7 were calculated as the weighted average of those of DIO and C1. The temperature dependence of the calculated coefficients for DIO, C1, and F7 is given in Fig. 3, Fig. 4, and Fig. 5, respectively. Considering the simplicity of the model, the calculated values show a good agreement with the measured ones.

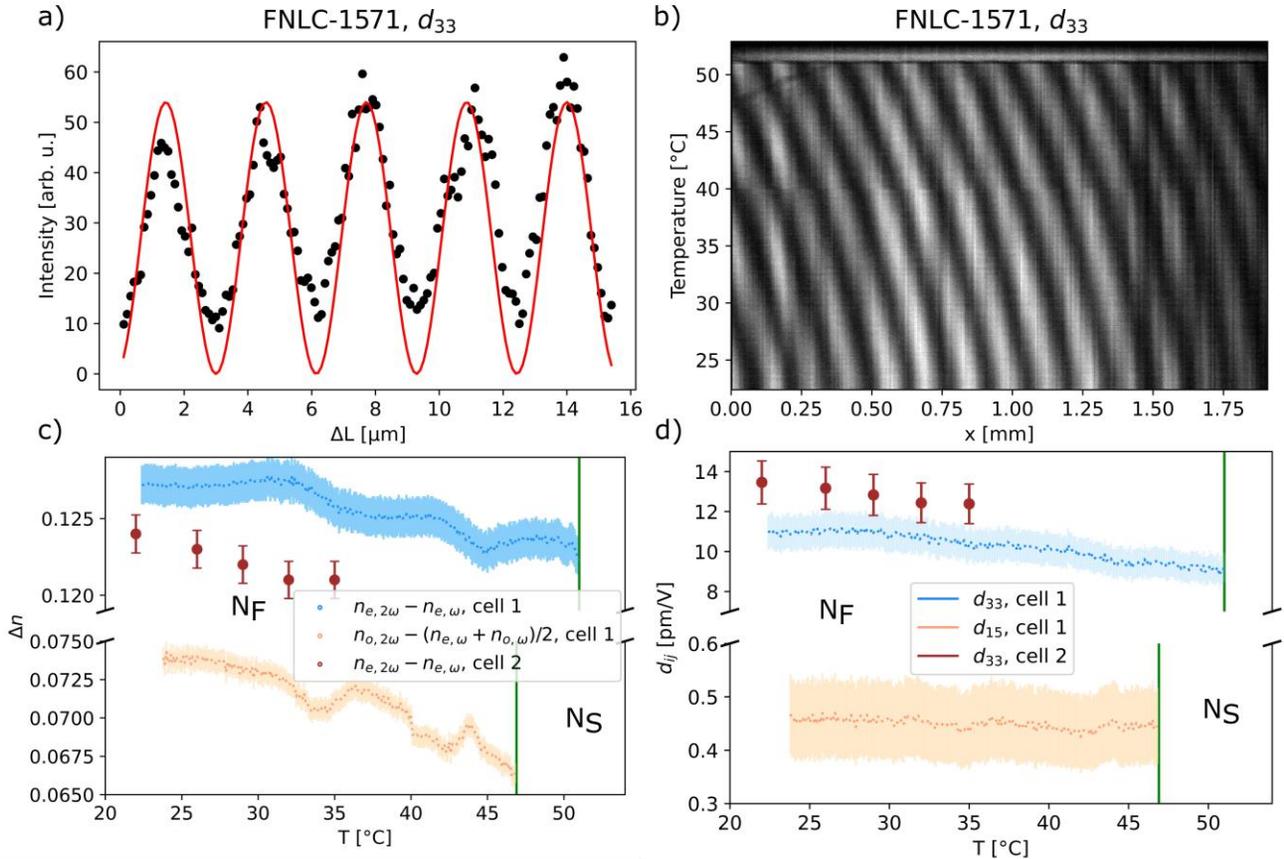

**Figure 6.** Second-order susceptibility measurement of FNLC-1571. a) SHG intensity as a function of cell thickness for FNLC-1571 (the measurement of d$_{33}$). b) The temperature dependence of the SHG intensity profile (the measurement of d$_{33}$). c) The measured refractive index mismatch Δn and d) the corresponding NLO susceptibility components d$_{33}$ and d$_{15}$ of FNLC-1571. The measured data (dots) with their uncertainties (shaded area). Cell 1 was the cell with a wedge slope of $\alpha = 1.27°$ where the Maker fringes could be analyzed from the image, while cell 2 was the cell with a smaller wedge slope of $\alpha = 0.07°$ that had to be translated for each measurement. The phase transition N$_F$-N$_S$ of FNLC-1571 occurred at 51°C during the measurement of d$_{33}$ and at 46.8°C during the measurement of d$_{15}$ (green line).

For the most accurate estimation of NLO coefficients, hyperpolarizability would need to be measured experimentally [33]. However, such measurements are delicate and time-consuming. With the increasing number of designed materials showing ferroelectric nematic and related phases, a thorough study of all to assess their potential for non-linear optics becomes an impractical task. Nevertheless, estimations obtained from DFT calculations show reasonably good agreement with experimental data. And while the full calculations are required for the most accurate estimation of the NLO properties, as polarization and order parameters of different FNLCs are usually on the same order of magnitude, a good indicator of high $d_{33}$ is given by the value of the hyperpolarizability component $\beta_{xxx}$, as is evident from the Tables S1-S3.

## 4. Conclusion

We have determined the NLO susceptibility values of several relevant FNLC materials, which show a wide dispersion of values, highlighting that, in addition to a material exhibiting the ferroelectric nematic phase, the molecular design is of vital importance for its potential for non-linear optics implementation. Our measurements show that the highest NLO coefficient value is still held by the first FNLC, RM734. However, this material was not designed or intended to have optimal NLO properties; there is no reason that a molecule or a mixture with better NLO properties could not be synthesized in the future. With additional optimization by the addition of chromophores [13], FNLCs could certainly surpass the NLO susceptibilities of the best solid-state NLO crystals.

A key challenge is then to be able to introduce effective π-bridged donor-acceptor groups with large electronic delocalization for maximizing NLO performance while at the same time preserving the molecular design requirements for retaining ferroelectric phases. Here, we have shown that molecular level hyperpolarizability calculations can be used for prescreening the potentiality of new materials. In combination with computational approaches that are being developed for the design and prediction of polar order in new molecular architectures [34], the presented approach can be used for assisted design of new materials with high NLO properties, accelerating the possibility of incorporating these new kind of polar materials into non-linear photonic applications.

## 5. Materials and Methods

### 5.1 Materials and Cells

Five liquid crystalline materials were investigated: RM734 [35], DIO [5], FNLC-1571 [20], C1 [24] and F7 [24]. RM734, DIO and FNLC-1571 and their phase sequences (I-N-$N_S$-$N_F$-crystal) are reported elsewhere [5,20,35,36]. The novel compound C1 is a pure compound which has a phase sequence Iso - 225.6 °C - N - 154.3 °C - SmA - 129.7 °C - SmA$_F$ - 90.1 °C - SmC$_P^H$. The material F7 is a mixture of 30% C1 and 70% DIO and has a phase sequence I – N – 97°C – N$_S$ — 86 °C – N$_F$ – 10°C – crystal. The materials RM734, DIO, C1, and F7 were synthesized according to the description given in references [24,35,37]. Structures of these materials are given in the SI (Supplementary Note I).

Two types of liquid crystal cells (assemblies in which liquid crystal material is sandwiched between two transparent substrates) were used. The first was the commercial wedge cells (KCRS-03, EHC) with surface coating rubbed parallel along the wedge and the wedge slope of $\alpha = 0.67°$. However, as some materials did not align well in these commercial cells, additional custom-made cells were prepared. These cells were composed of two indium tin oxide (ITO) coated glass plates, spin-coated with polyamide. Rubbing was applied in a parallel direction to the wedge. To create a wedge structure, spacers were placed only on one edge of the cell. The thickness of the cells was determined via spectral transmittance measurements (AvaSpec-2048, Avantes) and then additionally checked under the transmission microscope (Nikon Optiphot-2) using monochromatic light.

### 5.2. SHG setup

The fundamental light source for SHG measurements was a pulsed Ti:Sapphire laser (Coherent Legend) generating 50 fs long pulses with a repetition rate of 1 kHz. The initial 50 fs long pulses with a spectral width FWHM = 22 nm were extended to 200 fs with the use of a Fabry-Pérot filter. The resulting beam had a central wavelength of 800 nm and a spectral width FWHM = 1.5 nm. A beam splitter was used to divert a part of the beam to a power meter (Coherent LabMax_TOP) to monitor the average power during the measurements.

After passing through a $\lambda/2$ waveplate and the polarizer, the collimated linearly polarized fundamental beam impinged on the sample as depicted in Figure 1. A 5x Nikon objective collected the generated second harmonic beam to a CMOS camera (BFS-U3-17S7M-C, Blackfly, Teledyne FLIR). The typical exposure time was around 100 ms, the image size was 1600 x 1100 pixels, and the image resolution was 4.8 microns/pixel.

The temperature of the samples in commercial cells was controlled by an Instec HCS412W heating stage connected to an Instec mK2000 temperature controller. These cells have a wedge slope large enough that, within the laser beam spot size, multiple Maker fringes could be observed (Figure S6). For the custom-made cells with a smaller slope, the use of a translation stage was required to move the cell along the wedge to obtain a suitable thickness difference. These samples were heated using the Instec HCS302XY heating stage mounted on a translation stage, and the temperature was controlled with an Instec mK2000B temperature controller.

To measure a specific component of the nonlinear susceptibility tensor $d_{ij}$, the orientation of the fundamental beam polarization and the analyzer for the second harmonic beam were set as shown in Figure 1. In those cases for which no signal was obtained when using the Fabry-Pérot filter, which narrows the spectrum but reduces the optical power, the filter was removed from the optical path to increase the power. Without the filter, the spectrum of the fundamental beam exhibited quite a complex shape, as shown in Figure S3, which was also considered in the calculations.

The calibration of SHG measurements was performed by an NLO crystal sample, a Z-cut quartz. In this case, modification of the sample thickness and consequently Maker fringes were obtained by rotating the crystal plate around its vertical axis. A stepper motor rotation mount (K10CR1/M, Thorlabs) was used to control the rotation angle.

### 5.3 Maker Fringes Method

Since phase-matching could not be realized for any of the investigated materials, the nonlinear susceptibility tensor $d_{ij}$ values could not be directly measured [38]. Instead, the values were determined using the standard Maker Fringes method. The optical power of the second harmonic beam oscillates with the thickness of the SHG active material $L$ as [39–42]:

$$P_{2\omega} = \frac{S_{2\omega}}{S_\omega^2} \frac{\omega^2}{c_0^3 \varepsilon_0} \frac{1}{n_{2\omega}^2} T d_{eff}^2 P_\omega^2 \frac{\sin^2\left(\frac{2\pi L}{\lambda}\Delta n\right)}{\left(\frac{2\pi}{\lambda}\Delta n\right)^2}, \quad (9)$$

where $S_\omega$ and $S_{2\omega}$ are the cross-sectional areas of the fundamental and the SHG beam, $\omega$ is the fundamental beam frequency, $c_0$ the speed of light, $\varepsilon_0$ the vacuum permittivity, $n_{2\omega}$ the refractive index of the SHG active material at the frequency $2\omega$, $T$ is the Fresnel transmission factor, $d_{eff}$ is the effective nonlinear susceptibility, $P_\omega$ is the fundamental beam power, $\lambda$ is the fundamental beam wavelength, and $\Delta n = n_{2\omega} - n_\omega$ is the refractive index mismatch. The setup-dependent constants can be replaced by the parameter $\alpha$:

$$P_{2\omega} = \alpha \frac{1}{n_{2\omega}^2} T d_{eff}^2 P_\omega^2 \frac{\sin^2\left(\frac{2\pi L}{\lambda}\Delta n\right)}{\Delta n^2}. \quad (10)$$

In order to obtain the $d_{eff}$, a 500 μm-thick Z-cut quartz plate was used as a reference. The SHG signal from the quartz plate was measured while rotating it around the crystallographic X-axis, thus effectively changing the length of the SHG active medium. Comparing the maximal intensities obtained from the measured sample $P_{2\omega}^m$ and the reference quartz plate $P_{2\omega}^r$ the $d_{eff}$ could be calculated as:

$$d_{eff} = d_q \left(\frac{P_{2\omega}^m}{P_{2\omega}^r} \frac{T^r}{T^m}\right)^{\frac{1}{2}} \left(\frac{n_{2\omega}^m}{n_{2\omega}^r} * \frac{\Delta n^m}{\Delta n^r}\right). \quad (11)$$

The maximal intensities $P_{2\omega}^m$ were obtained by fitting. In the measurement without the Fabry-Pérot filter, the spectrum shown in Figure S3.b was considered. The optical parameters for quartz are well known, and their values are $d_{11} = 0.3 \frac{pm}{V}$ and $\Delta n = 0.0194$ [12]. For the calculation of the transmission factor of the liquid crystal cell $T^m$, all interfaces between air, glass, and liquid crystal were considered. The refractive index mismatch of the measured samples, $\Delta n^m$, was calculated from the period of the Maker fringes.

### 5.4 Birefringence and refractive index measurement

Birefringence ($\Delta n_{2\omega} = n_{2\omega,e} - n_{2\omega,o}$) of the FNLCs at the SHG wavelength was measured by placing the wedge cell between crossed polarizers in the transmission microscopy using monochromatic light and analyzing transmittivity as a function of thickness (Fig. S9). To assess the accuracy of the measurements of the refractive index mismatch, we compared the measured birefringence with the birefringence calculated from the values of refractive index mismatch, and the difference is below 2%, confirming the accuracy of the measurement.

The refractive index was measured by the optical interference method [43]. The wedge cell was observed under a microscope in reflection mode with an optical filter. To determine $n_e$ and $n_o$, light was polarized along and perpendicular to the director, respectively. The values of the refractive index were calculated from the period of the interference fringes.

### Declaration of Competing Interest

The authors declare no conflict of interest.

### Acknowledgements

The authors thank Janja Milivojević for carrying out the fabrication of the custom-made liquid crystal cells. The ferroelectric nematic liquid crystal FNLC-1571 used in this work was supplied by Merck Electronics KGaA. M.L., L.C., A.M, I.D.O., and N.S. acknowledge the support of the Slovenian Research Agency (grant numbers P1-0192, N1-0195, J1-50004, and PR-11214-1). C.J.G., J.H., and R.J.M. acknowledge funding from UKRI via a Future Leaders Fellowship, grant No. MR/W006391/1.

## Data availability

The data that support the findings of this study are available from the corresponding author upon reasonable request.

## References

[1] R. Dorn, D. Baums, P. Kersten, R. Regener, Nonlinear optical materials for integrated optics: Telecommunications and sensors, Advanced Materials 4 (1992) 464–473. https://doi.org/10.1002/adma.19920040703.

[2] F. De Martini, F. Sciarrino, Non-linear parametric processes in quantum information, Progress in Quantum Electronics 29 (2005) 165–256. https://doi.org/10.1016/j.pquantelec.2005.08.001.

[3] M.M. Fejer, Nonlinear Optical Frequency Conversion, Physics Today 47 (1994) 25–32. https://doi.org/10.1063/1.881430.

[4] J.V. Moloney, ed., Nonlinear Optical Materials, Springer, New York, NY, 1998. https://doi.org/10.1007/978-1-4612-1714-5.

[5] H. Nishikawa, K. Shiroshita, H. Higuchi, Y. Okumura, Y. Haseba, S. Yamamoto, K. Sago, H. Kikuchi, A Fluid Liquid-Crystal Material with Highly Polar Order, Advanced Materials 29 (2017) 1702354. https://doi.org/10.1002/adma.201702354.

[6] R. J. Mandle, S. J. Cowling, J. W. Goodby, A nematic to nematic transformation exhibited by a rod-like liquid crystal, Physical Chemistry Chemical Physics 19 (2017) 11429–11435. https://doi.org/10.1039/C7CP00456G.

[7] N. Sebastián, L. Cmok, R.J. Mandle, M.R. de la Fuente, I. Drevenšek Olenik, M. Čopič, A. Mertelj, Ferroelectric-Ferroelastic Phase Transition in a Nematic Liquid Crystal, Phys. Rev. Lett. 124 (2020) 037801. https://doi.org/10.1103/PhysRevLett.124.037801.

[8] X. Chen, E. Korblova, D. Dong, X. Wei, R. Shao, L. Radzihovsky, M.A. Glaser, J.E. Maclennan, D. Bedrov, D.M. Walba, N.A. Clark, First-principles experimental demonstration of ferroelectricity in a thermotropic nematic liquid crystal: Polar domains and striking electro-optics, PNAS 117 (2020) 14021–14031. https://doi.org/10.1073/pnas.2002290117.

[9] N. Sebastián, M. Čopič, A. Mertelj, Ferroelectric nematic liquid-crystalline phases, Phys. Rev. E 106 (2022) 021001. https://doi.org/10.1103/PhysRevE.106.021001.

[10] C.L. Folcia, J. Ortega, R. Vidal, T. Sierra, J. Etxebarria, The ferroelectric nematic phase: An optimum liquid crystal candidate for nonlinear optics, (n.d.) 22.

[11] R. Xia, X. Zhao, J. Li, H. Lei, Y. Song, W. Peng, X. Zhang, S. Aya, M. Huang, Achieving enhanced second-harmonic generation in ferroelectric nematics by doping D–π–A chromophores, J. Mater. Chem. C 11 (2023) 10905–10910. https://doi.org/10.1039/D3TC01384G.

[12] I. Shoji, T. Kondo, A. Kitamoto, M. Shirane, R. Ito, Absolute scale of second-order nonlinear-optical coefficients, J. Opt. Soc. Am. B 14 (1997) 2268. https://doi.org/10.1364/JOSAB.14.002268.

[13] J. Ortega, Folcia ,César Luis, T. and Sierra, High second harmonic generation in ferroelectric nematic liquid crystals by doping with optimally oriented chromophores, Liquid Crystals 0 (n.d.) 1–8. https://doi.org/10.1080/02678292.2025.2515115.

[14] H. Nishikawa, F. Araoka, A New Class of Chiral Nematic Phase with Helical Polar Order, Advanced Materials 33 (2021) 2101305. https://doi.org/10.1002/adma.202101305.

[15] X. Zhao, J. Zhou, J. Li, J. Kougo, Z. Wan, M. Huang, S. Aya, Spontaneous helielectric nematic liquid crystals: Electric analog to helimagnets, Proceedings of the National Academy of Sciences 118 (2021) e2111101118. https://doi.org/10.1073/pnas.2111101118.

[16] C. Feng, R. Saha, E. Korblova, D. Walba, S.N. Sprunt, A. Jákli, Electrically Tunable Reflection Color of Chiral Ferroelectric Nematic Liquid Crystals, Advanced Optical Materials 9 (2021) 2101230. https://doi.org/10.1002/adom.202101230.

[17] X. Zhao, H. Long, H. Xu, J. Kougo, R. Xia, J. Li, M. Huang, S. Aya, Nontrivial phase matching in helielectric polarization helices: Universal phase matching theory, validation, and electric switching, Proceedings of the National Academy of Sciences 119 (2022) e2205636119. https://doi.org/10.1073/pnas.2205636119.

[18] X. Zhao, J. Li, M. Huang, S. Aya, High-g-Factor Phase-Matched Circular Dichroism of Second Harmonic Generation in Chiral Polar Liquids, 2023.

[19] N. Sebastián, M. Lovšin, B. Berteloot, N. Osterman, A. Petelin, R.J. Mandle, S. Aya, M. Huang, I. Drevenšek-Olenik, K. Neyts, A. Mertelj, Polarization patterning in ferroelectric nematic liquids via flexoelectric coupling, Nat Commun 14 (2023) 3029. https://doi.org/10.1038/s41467-023-38749-2.

[20] M. Lovšin, A. Petelin, B. Berteloot, N. Osterman, S. Aya, M. Huang, I. Drevenšek-Olenik, R.J. Mandle, K. Neyts, A. Mertelj, N. Sebastian, Patterning of 2D second harmonic generation active arrays in ferroelectric nematic fluids, Giant 19 (2024) 100315. https://doi.org/10.1016/j.giant.2024.100315.

[21] N. Sebastián, R.J. Mandle, A. Petelin, A. Eremin, A. Mertelj, Electrooptics of mm-scale polar domains in the ferroelectric nematic phase, Liquid Crystals 0 (2021) 1–17. https://doi.org/10.1080/02678292.2021.1955417.

[22] V. Sultanov, A. Kavčič, E. Kokkinakis, N. Sebastián, M.V. Chekhova, M. Humar, Tunable entangled photon-pair generation in a liquid crystal, Nature (2024) 1–6. https://doi.org/10.1038/s41586-024-07543-5.

[23] J. Karcz, J. Herman, N. Rychłowicz, P. Kula, E. Górecka, J. Szydlowska, P.W. Majewski, D. Pociecha, Spontaneous chiral symmetry breaking in polar fluid-heliconical ferroelectric nematic phase, Science 384 (2024) 1096–1099. https://doi.org/10.1126/science.adn6812.

[24] C.J. Gibb, J. Hobbs, D.I. Nikolova, T. Raistrick, S.R. Berrow, A. Mertelj, N. Osterman, N. Sebastián, H.F. Gleeson, R.J. Mandle, Spontaneous symmetry breaking in polar fluids, Nat Commun 15 (2024) 5845. https://doi.org/10.1038/s41467-024-50230-2.

[25] A. Erkoreka, M. Huang, S. Aya, J. Martinez-Perdiguero, Mean-Field Theory of the Uniaxial Ferroelectric Smectic A Liquid Crystal Phase, Phys. Rev. Lett. 133 (2024) 208101. https://doi.org/10.1103/PhysRevLett.133.208101.

[26] H. Kikuchi, H. Matsukizono, K. Iwamatsu, S. Endo, S. Anan, Y. Okumura, Fluid Layered Ferroelectrics with Global C∞v Symmetry, Advanced Science 9 (2022) 2202048. https://doi.org/10.1002/advs.202202048.

[27] D.A. Dunmur, Local field effects in oriented nematic liquid crystals, Chemical Physics Letters 10 (1971) 49–51. https://doi.org/10.1016/0009-2614(71)80153-7.

[28] A.A. Minko, Rachkevich ,V. S., S.Ye. and Yakovenko, Invited Article. The local field in uniaxial liquid crystals, Liquid Crystals 4 (1989) 1–19. https://doi.org/10.1080/02678298908028955.

[29] P. Palffy-Muhoray, D.A. Balzarini, Refractive index measurements and order parameter determination of the liquid crystal p-ethoxybenzilidene-p-n-butylaniline, Can. J. Phys. 59 (1981) 515–520. https://doi.org/10.1139/p81-066.

[30] A. Erkoreka, N. Sebastián, A. Mertelj, J. Martinez-Perdiguero, A molecular perspective on the emergence of long-range polar order from an isotropic fluid, Journal of Molecular Liquids 407 (2024) 125188. https://doi.org/10.1016/j.molliq.2024.125188.

[31] P. Günter, ed., Nonlinear Optical Effects and Materials, Springer, Berlin, Heidelberg, 2000. https://doi.org/10.1007/978-3-540-49713-4.

[32] L.T. Cheng, W. Tam, S.H. Stevenson, G.R. Meredith, G. Rikken, S.R. Marder, Experimental investigations of organic molecular nonlinear optical polarizabilities. 1. Methods and results on benzene and stilbene derivatives, J. Phys. Chem. 95 (1991) 10631–10643. https://doi.org/10.1021/j100179a026.

[33] F. Araoka, B. Park, Y. Kinoshita, H. Takezoe, J. Thisayukta, J. Watanabe, Determination of Nonlinear Hyperpolarizability of Bent-Shaped Molecules using Hyper-Rayleigh Scattering, Molecular Crystals and Liquid Crystals Science and Technology. Section A. Molecular Crystals and Liquid Crystals 328 (1999) 291–297. https://doi.org/10.1080/10587259908026070.

[34] J. Hobbs, C.J. Gibb, R.J. Mandle, Polarity from the bottom up: a computational framework for predicting spontaneous polar order, J. Mater. Chem. C 13 (2025) 13367–13375. https://doi.org/10.1039/D5TC01641J.

[35] R.J. Mandle, S.J. Cowling, J.W. Goodby, Rational Design of Rod-Like Liquid Crystals Exhibiting Two Nematic Phases, Chemistry – A European Journal 23 (2017) 14554–14562. https://doi.org/10.1002/chem.201702742.

[36] J. Thoen, G. Cordoyiannis, W. Jiang, G.H. Mehl, C. Glorieux, Phase transitions study of the liquid crystal DIO with a ferroelectric nematic, a nematic, and an intermediate phase and of mixtures with the ferroelectric nematic compound RM734 by adiabatic scanning calorimetry, Phys. Rev. E 107 (2023) 014701. https://doi.org/10.1103/PhysRevE.107.014701.

[37] J. Li, H. Nishikawa, J. Kougo, J. Zhou, S. Dai, W. Tang, X. Zhao, Y. Hisai, M. Huang, S. Aya, Development of ferroelectric nematic fluids with giant-ε dielectricity and nonlinear optical properties, Science Advances 7 (n.d.) eabf5047. https://doi.org/10.1126/sciadv.abf5047.

[38] G.E. Francois, cw Measurement of the Optical Nonlinearity of Ammonium Dihydrogen Phosphate, Phys. Rev. 143 (1966) 597–600. https://doi.org/10.1103/PhysRev.143.597.

[39] J. Jerphagnon, S.K. Kurtz, Maker Fringes: A Detailed Comparison of Theory and Experiment for Isotropic and Uniaxial Crystals, Journal of Applied Physics 41 (1970) 1667–1681. https://doi.org/10.1063/1.1659090.

[40] A. Majkić, A. Franke, R. Kirste, R. Schlesser, R. Collazo, Z. Sitar, M. Zgonik, Optical nonlinear and electro-optical coefficients in bulk aluminium nitride single crystals, Physica Status Solidi (b) 254 (2017) 1700077. https://doi.org/10.1002/pssb.201700077.

[41] W.N. Herman, L.M. Hayden, Maker fringes revisited: second-harmonic generation from birefringent or absorbing materials, J. Opt. Soc. Am. B 12 (1995) 416. https://doi.org/10.1364/JOSAB.12.000416.

[42] P.D. Maker, R.W. Terhune, M. Nisenoff, C.M. Savage, Effects of Dispersion and Focusing on the Production of Optical Harmonics, Phys. Rev. Lett. 8 (1962) 21–22. https://doi.org/10.1103/PhysRevLett.8.21.

[43] P. Kumari, B. Basnet, M.O. Lavrentovich, O.D. Lavrentovich, Chiral ground states of ferroelectric liquid crystals, Science 383 (2024) 1364–1368. https://doi.org/10.1126/science.adl0834.

# Supplementary Material

## Ferroelectric Fluids for Nonlinear Photonics: Evaluation of Temperature Dependence of Second-Order Susceptibilities

M. Lovšin[1,2], L. Cmok[1], C. J. Gibb[3], J. Hobbs[4], R. J. Mandle[3,4], A. Mertelj[1], I. Drevenšek-Olenik[1,2], N. Sebastian[1]

1 Jožef Stefan Institute, Ljubljana, Slovenia
2 University of Ljubljana, Faculty of Mathematics and Physics, Ljubljana, Slovenia
3 School of Chemistry, University of Leeds, Leeds, UK
4 School of Physics and Astronomy, University of Leeds, Leeds, UK

**Index**

# Supplementary Note I: Molecular Structures of Materials Used

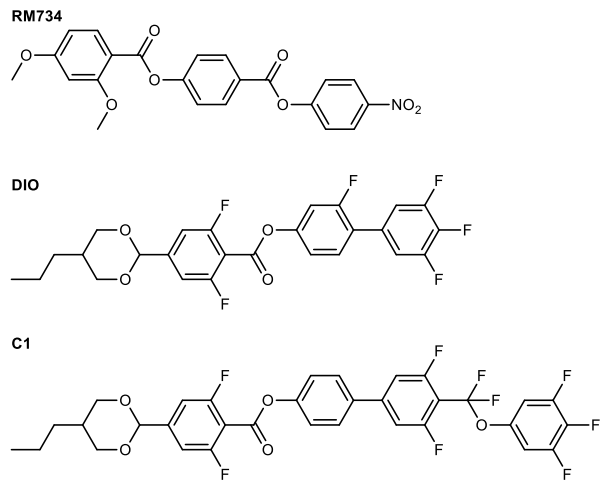

Figure S1: Molecular structures of materials used in this study. The mixture F7 is a binary mixture of DIO (70 mol%) and C1 (30 mol%).

# Supplementary Note II: SHG setup

As the sample of lithium niobate has larger refractive index and absorption than quartz at the wavelength 800 nm (Fig. S1), the measurement error is larger, and the quartz sample was decided as more suitable to be used as a reference for the measurements of NLO susceptibilities of ferroelectric nematic liquid crystals (FNLCs).

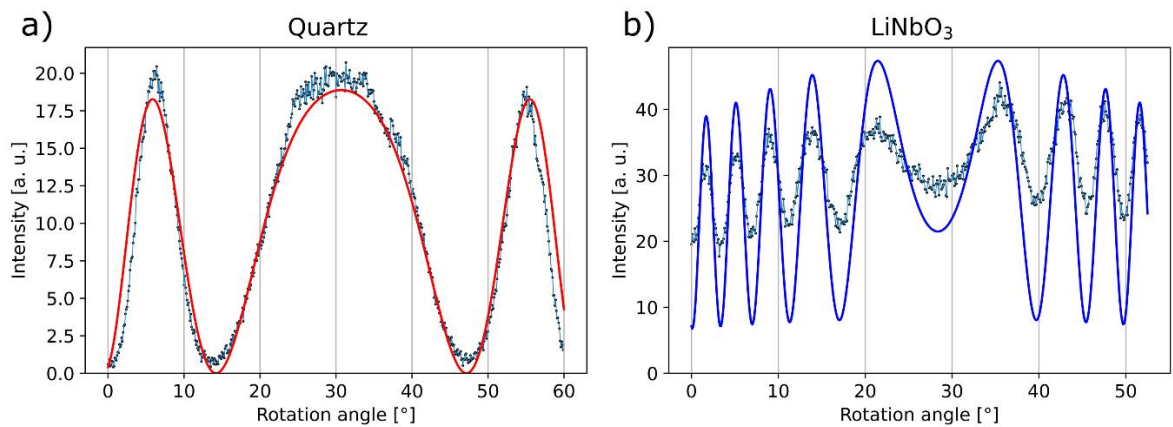

Figure S2: a) Measurement (black dots) and fitted curve (red line) of the SHG signal of the reference sample of quartz rotated around the vertical axis, and b) the measurement (black dots) and fitted curve (blue line) of the lithium niobate.

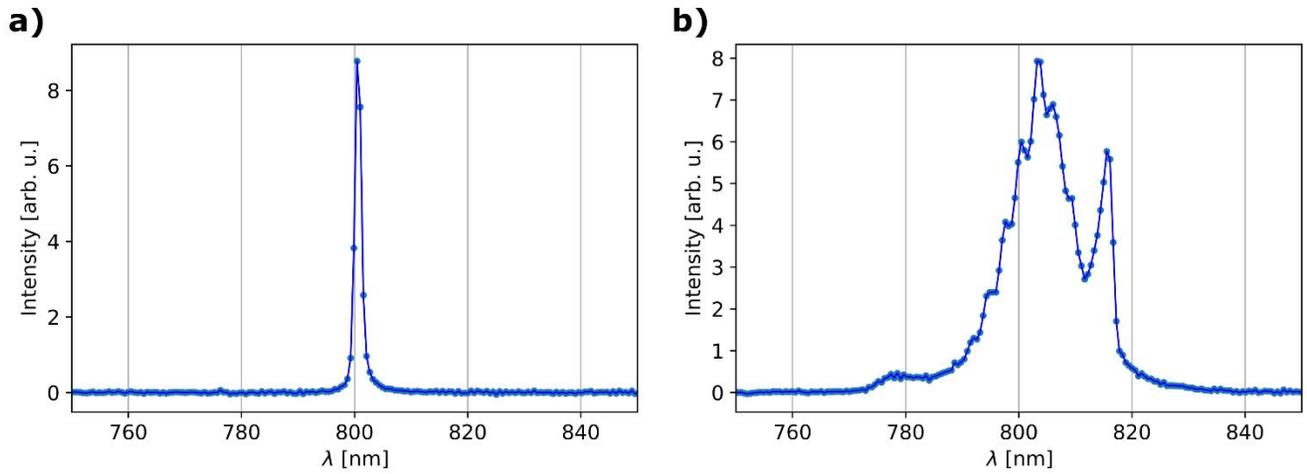

Figure S3: Spectrum of the pumping laser a) with and b) without the Fabry-Pérot filter.

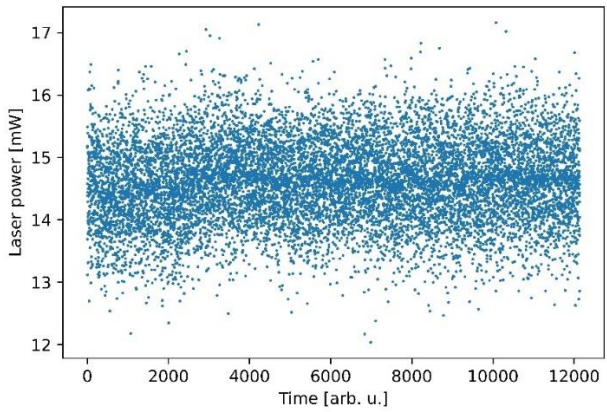

Figure S4: Average laser power during the measurement of RM734.

## Supplementary Note III: Additional images

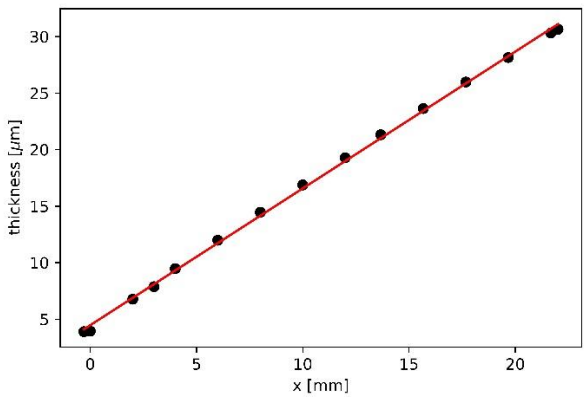

Figure S5: Thickness profile of the custom-made cell used for DIO. The thickness was determined by analysis of the optical transmittance spectrum of the cell.

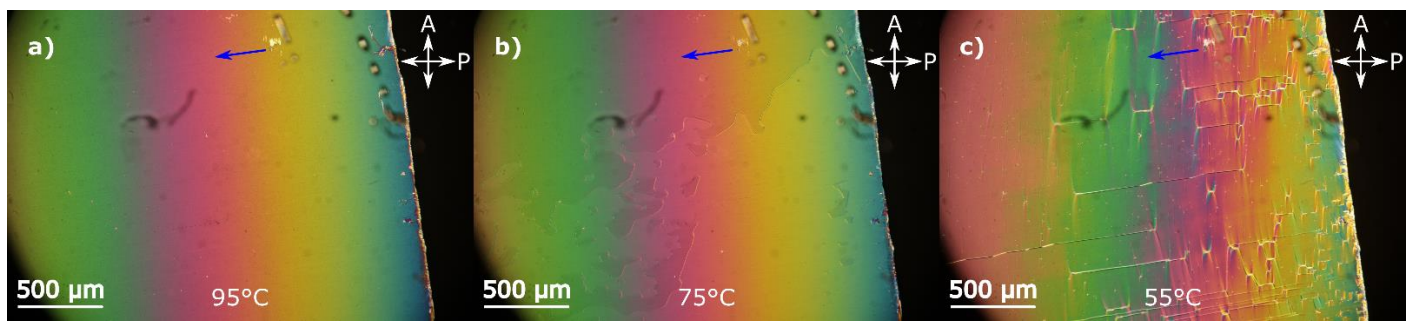

Figure S6: POM images of DIO in a custom-made wedge cell at different phases: a) the N phase, b) $N_s$ phase, and c) $N_F$ phase. The polarizers are in extinction position (white arrows). The cell is at an angle of 8°, and its rubbing is along the wedge (blue arrow).

## Supplementary Note IV: SHG fringes images

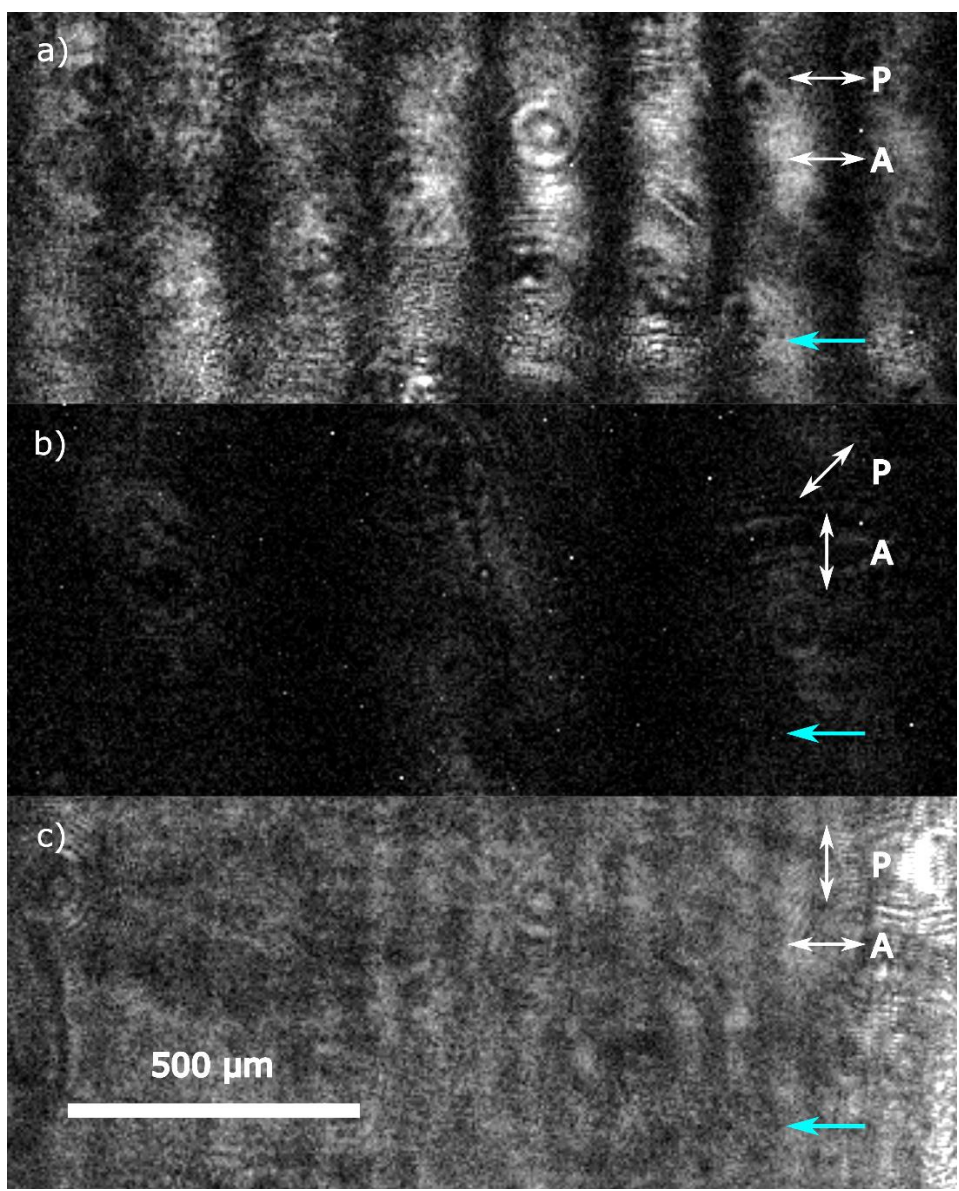

Figure S7: SHG images of Maker fringes in the EHC wedge cell filled with RM734. The wedge in the images is oriented horizontally. Measurement of a) d33, b) d15, and c) d31. The measurements (a) and (b) are performed with the Fabry-Pérot filter and with exposure times of 250 ms (a) and 100 ms (b) while the measurement (c) is performed without the Fabry-Pérot filter and with an exposure time of 10 ms. The brightness of all three images above was increased by 30% for better clarity.

## Supplementary Note V: Local field factor

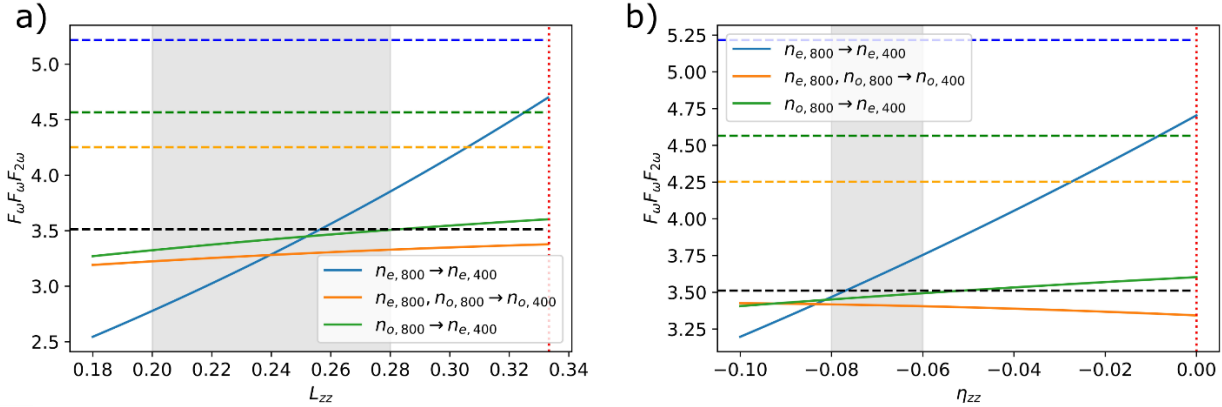

Figure S8: The local field factors for the calculations of d33 (full blue line), d15 (full orange line), and d31 (full green line) calculated using the a) anisotropic crystal model [1] and b) anisotropic liquid model [2] compared to the calculations considering dispersion, $F(\omega) = \frac{n^2+2}{3}$, (dashed lines with corresponding color) and the calculation with the average $n = \frac{1}{3}(2n_{o,\omega} + n_{e,\omega})$ (dashed black line). The red dotted line marks the isotropic case (at $L_{ee} = \frac{1}{3}$ and $\eta_{ee} = 0$). The typical values for nematic liquid crystals are $L_{ee}$ between 0.2 and 0.28 [2] and $\eta_{ee}$ between 0.06 and 0.08 [3] (highlighted grey area).

## Supplementary Note VI: Additional information for $d_{ij}$ calculations

The thermally averaged hyperpolarizability components in the laboratory frame can be calculated from the ones in the molecular frame according to [4]:

$$\langle \beta_{33} \rangle = \frac{1}{5} \left[ (\langle P_1 \rangle - \langle P_3 \rangle)(2\beta_{yyx} + 2\beta_{zzx} + \beta_{xyy} + \beta_{xzz}) + (3\langle P_1 \rangle + 3\langle P_3 \rangle)\beta_{xxx} \right] \quad \text{(S1)}$$

$$\langle \beta_{15} \rangle = \frac{1}{10} \left[ \langle P_3 \rangle (2\beta_{yyx} + 2\beta_{zzx} + \beta_{xyy} + \beta_{xzz} - 2\beta_{xxx}) + \langle P_1 \rangle (3\beta_{yyx} + 3\beta_{zzx} - \beta_{xyy} - \beta_{xzz} + 2\beta_{xxx}) \right] \quad \text{(S2)}$$

$$\langle \beta_{31} \rangle = \frac{1}{10} \left[ \langle P_3 \rangle (2\beta_{yyx} + 2\beta_{zzx} + \beta_{xyy} + \beta_{xzz} - 2\beta_{xxx}) - 2\langle P_1 \rangle (\beta_{yyx} + \beta_{zzx} - 2\beta_{xyy} - 2\beta_{xzz} - \beta_{xxx}) \right] \quad \text{(S3)}$$

The components of the hyperpolarizability tensor in the molecular frame were obtained by calculating the molecular electronic structure with the Gaussian G09 rev D01 at the M06HF-D3/aug-cc-pVTZ level at 800 nm. Gaussian output hyperpolarizability values for RM734, DIO, and C1 are shown in Tables S1-S3. One should note here that the final used values enter equations S1-S3 multiplied by the factor $1/(4\varepsilon_0)$, where the factor ¼ comes from the 1/n! with n=2 factor not accounted by Gaussian software and an additional ½ factor coming from the conversion between the experimental and standard different hyperpolarizabilities definitions as thoroughly discussed in Chapter 2, in [5].

| $\beta_{xxx}$ | −12.45 | $\beta_{yyy}$ | −0.05 | $\beta_{zzz}$ | 0.00 |
|---|---|---|---|---|---|
| $\beta_{xyx}$ | −2.60 | $\beta_{yyx}$ | 0.45 | $\beta_{zzx}$ | 0.28 |
| $\beta_{xyy}$ | 0.60 | $\beta_{yxx}$ | −2.30 | $\beta_{zxx}$ | 0.20 |
| $\beta_{xzx}$ | 0.22 | $\beta_{yzy}$ | 0.00 | $\beta_{zzy}$ | 0.03 |
| **$\beta_{xzz}$** | **0.30** | $\beta_{yzz}$ | 0.04 | $\beta_{zyy}$ | 0.00 |
| $\beta_{xzy}$ | 0.01 | $\beta_{yzx}$ | 0.01 | $\beta_{zyx}$ | 0.01 |

Table S1: First-order hyperpolarizability values of RM734 in SI units $\left(*10^{-50}\frac{C^3 m^3}{J^2}\right)$ calculated by Gaussian. The molecular dipole is along the x-axis. The components that enter Equations S1-S3 are highlighted green.

| | | | | | |
|---|---|---|---|---|---|
| **$\beta_{xxx}$** | **−1.30** | $\beta_{yyy}$ | −0.09 | $\beta_{zzz}$ | −0.06 |
| $\beta_{xyx}$ | 0.12 | **$\beta_{yyx}$** | **0.22** | **$\beta_{zzx}$** | **0.19** |
| **$\beta_{xyy}$** | **0.24** | $\beta_{yxx}$ | 0.11 | $\beta_{zxx}$ | −0.05 |
| $\beta_{xzx}$ | −0.06 | $\beta_{yzy}$ | 0.04 | $\beta_{zzy}$ | −0.01 |
| **$\beta_{xzz}$** | **0.20** | $\beta_{yzz}$ | −0.01 | $\beta_{zyy}$ | 0.04 |
| $\beta_{xzy}$ | 0.01 | $\beta_{yzx}$ | 0.01 | $\beta_{zyx}$ | 0.01 |

Table S2: First-order hyperpolarizability values of C1 in SI units $\left(*10^{-50}\frac{C^3 m^3}{J^2}\right)$ calculated by Gaussian. The molecular dipole is along the x-axis. The components that enter Equations S1-S3 are highlighted green.

| | | | | | |
|---|---|---|---|---|---|
| **$\beta_{xxx}$** | **−0.30** | $\beta_{yyy}$ | −0.10 | $\beta_{zzz}$ | −0.04 |
| $\beta_{xyx}$ | 0.12 | **$\beta_{yyx}$** | **0.21** | **$\beta_{zzx}$** | **0.09** |
| **$\beta_{xyy}$** | **0.22** | $\beta_{yxx}$ | 0.10 | $\beta_{zxx}$ | 0.00 |
| $\beta_{xzx}$ | −0.01 | $\beta_{yzy}$ | −0.04 | $\beta_{zzy}$ | 0.01 |
| **$\beta_{xzz}$** | **0.10** | $\beta_{yzz}$ | 0.01 | $\beta_{zyy}$ | −0.04 |
| $\beta_{xzy}$ | −0.01 | $\beta_{yzx}$ | −0.01 | $\beta_{zyx}$ | −0.01 |

Table S3: First-order hyperpolarizability values of DIO in SI units $\left(*10^{-50}\frac{C^3 m^3}{J^2}\right)$ calculated by Gaussian. The molecular dipole is along the x-axis. The components that enter Equations S1-S3 are highlighted green.

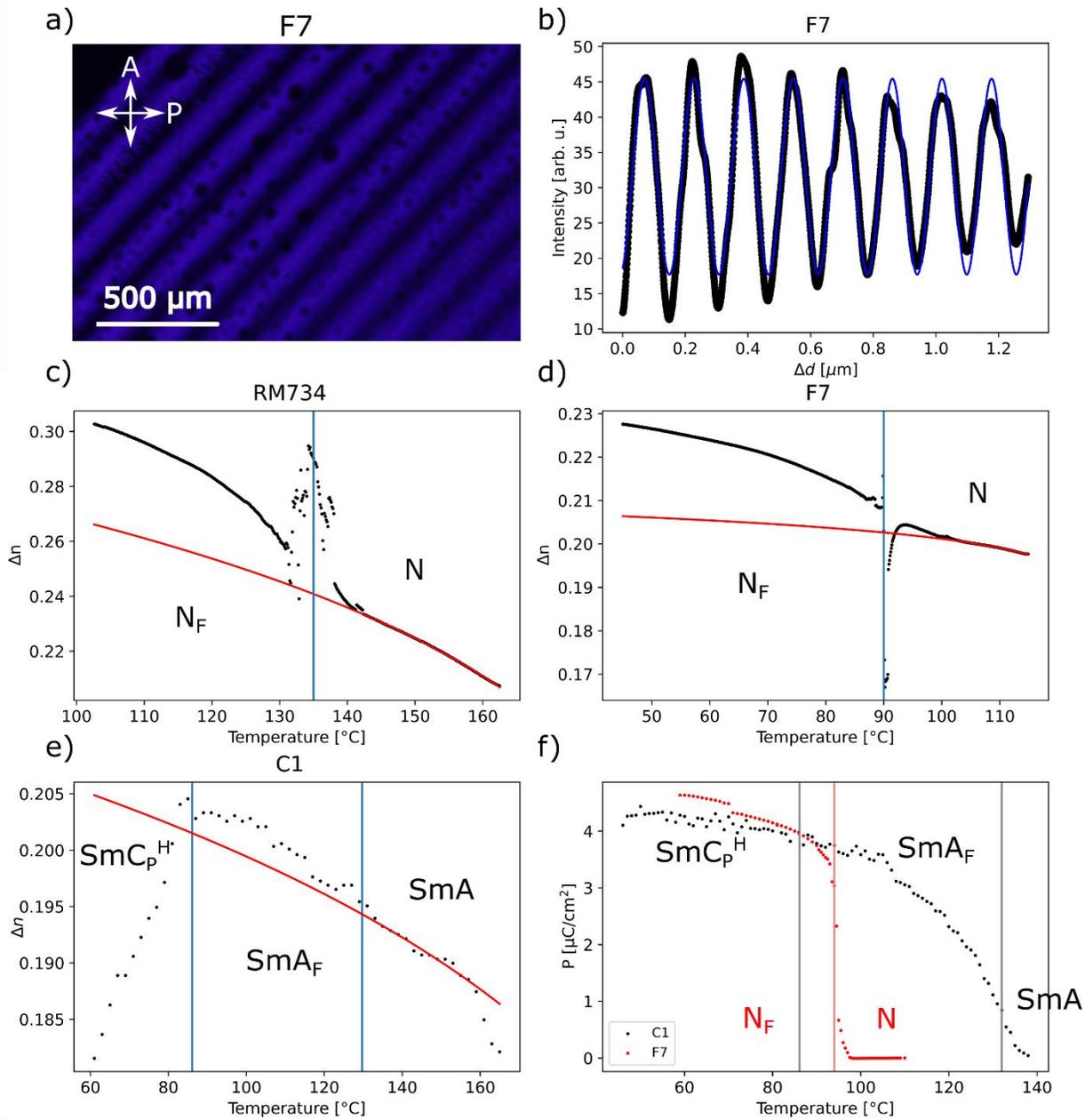

Figure S9: a) A sample of F7 in a wedge cell under crossed polarizers, illuminated with monochromatic (405 nm) light. b) The transmitted intensity vs cell thickness. The calculated birefringence (black dots) of c) RM734, d) F7, and e) C1. The red line is the extrapolation of the birefringence from the N phase (for C1 from the SmA phase). The ratio between the birefringence in the $N_F$ phase (for C1 in the $SmA_F$ phase) and the extrapolated birefringence is used in Equation (12). f) The temperature dependence of polarization of C1 [6] and F7.

|  | $n_o$ | $n_e$ | ρ [kg/m³] | M [g/mol] | μ [D] |
|---|---|---|---|---|---|
| RM734 | 1.5 | 1.8 | 1280 | 424 | 11.4 |
| DIO | 1.44 | 1.74 | 1300 | 510.39 | 9.4 |
| C1 | 1.5 | 1.6 | 1300 | 670.47 | 11.9 |
| F7 | 1.42 | 1.62 | 1300 | 558.42 | 10.15 |

Table S4: Values used for $d_{ij}$ calculations: ordinary and extraordinary refractive index, density, molar mass, and dipole moment of RM734, DIO, C1, and F7. Molar mass and dipole moment of F7 were calculated as the weighted average of DIO and C1. The values of the dipole moment μ of RM734 and DIO were taken from [7] and [8].

## Supplementary Note VII: Comparison between methods

We have additionally filled the FNLC-1571 in a liquid crystal cell with a small wedge slope of around $tan\theta = 0.001$ to compare the methods of analyzing the Maker fringes from the image or by translation of the sample. Both methods give comparable values of $d_{33}$.

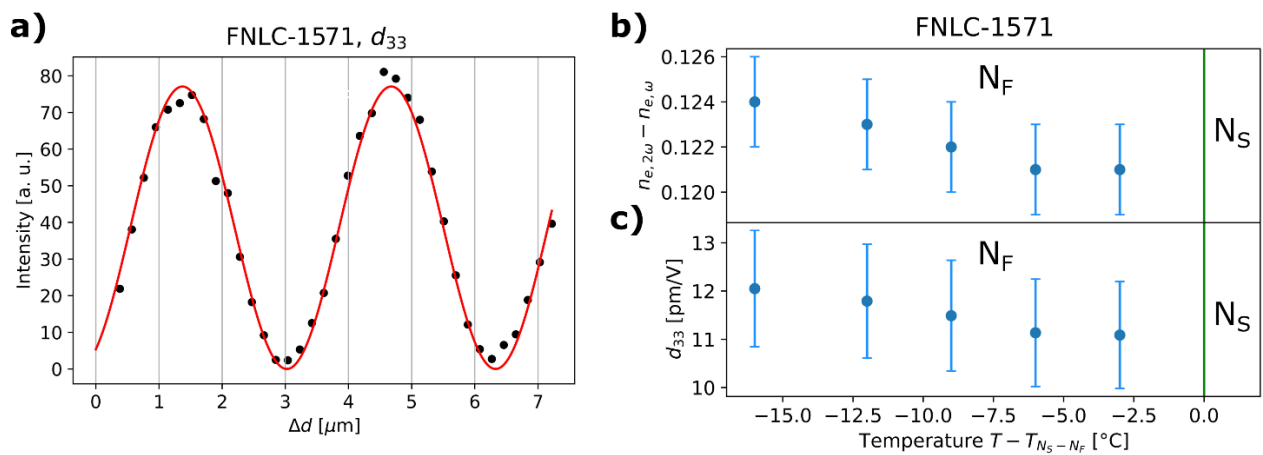

Figure S10: Second-order susceptibility measurement of FNLC-1571 in a wedge cell with a smaller wedge slope. a) SHG intensity as a function of cell thickness for FNLC-1571 (the measurement of $d_{33}$). b) The measured Δn and c) the corresponding NLO coefficient components $d_{33}$ of FNLC-1571.

## References

[1] D.A. Dunmur, Local field effects in oriented nematic liquid crystals, Chemical Physics Letters 10 (1971) 49–51. https://doi.org/10.1016/0009-2614(71)80153-7.
[2] A.A. Minko, Rachkevich ,V. S., S.Ye. and Yakovenko, Invited Article. The local field in uniaxial liquid crystals, Liquid Crystals 4 (1989) 1–19. https://doi.org/10.1080/02678298908028955.
[3] P. Palffy-Muhoray, D.A. Balzarini, Refractive index measurements and order parameter determination of the liquid crystal p-ethoxybenzilidene-p-n-butylaniline, Can. J. Phys. 59 (1981) 515–520. https://doi.org/10.1139/p81-066.
[4] A. Erkoreka, N. Sebastián, A. Mertelj, J. Martinez-Perdiguero, A molecular perspective on the emergence of long-range polar order from an isotropic fluid, Journal of Molecular Liquids 407 (2024) 125188. https://doi.org/10.1016/j.molliq.2024.125188.

[5] P. Günter, ed., Nonlinear Optical Effects and Materials, Springer, Berlin, Heidelberg, 2000. https://doi.org/10.1007/978-3-540-49713-4.

[6] C.J. Gibb, J. Hobbs, D.I. Nikolova, T. Raistrick, S.R. Berrow, A. Mertelj, N. Osterman, N. Sebastián, H.F. Gleeson, R.J. Mandle, Spontaneous symmetry breaking in polar fluids, Nat Commun 15 (2024) 5845. https://doi.org/10.1038/s41467-024-50230-2.

[7] R.J. Mandle, N. Sebastián, J. Martinez-Perdiguero, A. Mertelj, On the molecular origins of the ferroelectric splay nematic phase, Nat Commun 12 (2021) 4962. https://doi.org/10.1038/s41467-021-25231-0.

[8] H. Nishikawa, K. Shiroshita, H. Higuchi, Y. Okumura, Y. Haseba, S. Yamamoto, K. Sago, H. Kikuchi, A Fluid Liquid-Crystal Material with Highly Polar Order, Advanced Materials 29 (2017) 1702354. https://doi.org/10.1002/adma.201702354.